\documentclass{sig-alternate-2013}

\usepackage{comment}
\usepackage{subfig}
\usepackage[font={small,bf}]{caption}
\usepackage{fixltx2e} 

\usepackage{inconsolata}

\DeclareRobustCommand{\ttfamily}{\fontfamily{lmtt}\selectfont}

\newcommand{\code}[1]{{\small\protect\texttt{\vspace{-5pt}\newline#1}}\vspace{-5pt}\newline}
\newcommand{\tab}[1]{\noindent\hspace*{#1cm}}
\newcommand{\ntab}[1]{\newline\noindent\hspace*{#1cm}}
\newcommand{\ic}[1]{\texttt{\small{#1}}}

\usepackage{balance}  
\usepackage{graphics} 
\usepackage{times}    

\def\pprw{8.5in}
\def\pprh{11in}

\permission{Copyright is held by the International World Wide Web Conference Committee (IW3C2). IW3C2 reserves the right to provide a hyperlink to the author's site if the Material is used in electronic media.}
\conferenceinfo{WWW'14,}{April 7--11, 2014, Seoul, Korea.} 
\copyrightetc{ACM \the\acmcopyr}
\crdata{978-1-4503-2744-2/14/04. \\
http://dx.doi.org/10.1145/2566486.2567967}

\newcommand\tabhead[1]{\small\textbf{#1}}

\begin{document}

\title{Designing and Deploying Online Field Experiments}

\numberofauthors{3}
\author{
  \alignauthor Eytan Bakshy\\
    \affaddr{Facebook}\\
    \affaddr{Menlo Park, CA}\\
    \affaddr{eytan@fb.com}\\
  \alignauthor Dean Eckles\\
    \affaddr{Facebook}\\
    \affaddr{Menlo Park, CA}\\
    \affaddr{deaneckles@fb.com}\\
  \alignauthor Michael S. Bernstein\\
    \affaddr{Stanford University}\\
    \affaddr{Palo Alto, CA}\\
    \affaddr{msb@cs.stanford.edu}\\
}
\maketitle

\begin{abstract}

Online experiments are widely used to compare specific design alternatives, but they can also be used to produce generalizable knowledge and inform strategic decision making. Doing so often requires sophisticated experimental designs, iterative refinement, and careful logging and analysis. Few tools exist that support these needs. We thus introduce a language for online field experiments called \emph{PlanOut}. PlanOut separates experimental design from application code, allowing the experimenter to concisely describe experimental designs, whether common ``A/B tests'' and factorial designs, or more complex designs involving conditional logic or multiple experimental units. These latter designs are often useful for understanding causal mechanisms involved in user behaviors. We demonstrate how experiments from the literature can be implemented in PlanOut, and describe two large field experiments conducted on Facebook with PlanOut. For common scenarios in which experiments are run iteratively and in parallel, we introduce a namespaced management system that encourages sound experimental practice.

\end{abstract}


\category{H.5.3.}{Group and Organization Interfaces}{Evaluation / methodology}

\section{Introduction}
Randomized field experiments are central to contemporary design and development processes for Internet services. In the most popular case, practitioners use ``A/B tests'' that randomly assign users to one of two variations of a service. Doing so often allows designers and developers to quickly identify the best choice of the two.
The Internet industry has distinct advantages in how organizations can use experiments to make decisions: developers can introduce numerous variations on the service without substantial engineering or distribution costs, and observe how a large random sample of users (rather than a convenience sample) behave when randomly assigned to these variations. So, in many ways, experimentation with Internet services is easy. 

For some organizations, randomized experiments play a central role throughout the design and decision-making process.  Experiments may be used to explore a design space \cite{kohavi2009controlled}, better attribute outcomes to causes  \cite{bakshy2012social,farahat2012effective}, and estimate effects that help decision makers understand how people react to changes and use their services \cite{manzi2012uncontrolled,watts2011everything}. In this way, even early stages of the user-centered design process can be informed by field experiments.  Such trials are not intended to optimize short-term objectives through a pick-the-winner process; they instead aim to provide more lasting, generalizable knowledge.  While experiments that achieve these objectives often draw from experimental designs used in the behavioral and social sciences, available tools do little to support the design, deployment, or analysis of these more sophisticated experiments.

The realities surrounding the deployment of routine experiments can make their evaluation quite complex.
Online experimentation is highly iterative, such that preliminary results are used to rapidly run follow-up experiments. This necessitates changing or launching new experiments. Changing live experiments can easily result in statistical inferences that are incorrect. From a development perspective, running follow-up experiments can be time consuming and error prone because experimental logic is often mixed in with application code.
Online experimentation is also distributed across individuals and teams, and over time. This can make it difficult to run experiments simultaneously without interacting with other's experiments or complicating application logic.
Combined, these features make it so that experimentation can become so embedded in application code that only a few engineers can correctly modify a particular experiment without introducing errors in its design or future analysis.

In this work, we discuss how Internet-scale field experiments can be designed and deployed with \emph{PlanOut}, a domain-specific language for experimentation used at Facebook.
Designers and engineers working with PlanOut can view any aspect of a service as tunable via parameters: e.g., a flag signaling whether a banner is visible, a variable encoding the number of items in an aggregation, or a string that corresponds to the text of a button. Experimental logic is encapsulated in simple scripts that assign values to parameters. Basic random assignment primitives can be combined to reliably implement complex experimental designs, including those that involve assignment of multiple experimental units to multiple factors and treatments with continuous values.  PlanOut scripts can also be used as a concise summary of an experiment's design and manipulations, which makes it easier to communicate about, and replicate experiments. 

PlanOut is enough by itself to author one-off, isolated experiments quickly. However, if experiments must be iterated upon, or related experiments must run in parallel, additional infrastructure is necessary. Such systems can be used to manage experiments and logging, prevent interference between experiments. In this paper, we introduce the architecture of a management system for such situations, and illustrate how iterative experiments can soundly run and analyzed.

In sum, this paper contributes:
\begin{itemize}
  \item A characterization of online experimentation based on parametrization of user experiences,
  \item The PlanOut language, which provides a high-ceiling, low-threshold toolkit for parameter-based experiments,
  \item Guidelines for managing and analyzing iterative and distributed experiments.

\end{itemize}
\noindent
These contributions together articulate a perspective on how online field experiments should be conceptualized and implemented. This perspective advocates the use of short, centralized scripts to describe assignment procedures at a high level. It results in experimental practice that is more agile and encourages the production of generalizable, scientific knowledge.

The paper is structured as follows.  After reviewing related work in Section~\ref{sec:related}, we introduce the PlanOut language in Section~\ref{sec:planout} and show how it can be used to design both simple and complex experiments.  Then, in Sections~\ref{sec:running} and Section~\ref{sec:Analysis}, we describe how distributed, iterative experiments can be managed, logged, and analyzed. Finally, we discuss the broader implications and and limitations of our work in Section~\ref{sec:discussion}.
\section{Related Work}
\label{sec:related}
The design and analysis of experiments is a developed area within statistics that is regularly taught, with domain-specific elements, to students in industrial engineering, psychology, marketing, human--computer interaction, and other fields \cite{box2005statistics,gerber2012field,manzi2012uncontrolled,shadish_renaissance_2009}. Tools for producing design matrices for factorial and fractional factorial designs are available (e.g., in the R packages \ic{DoE}, \ic{Design}, and \ic{rms}).  Such packages are useful for designing small scale studies, but since the design matrix is created \emph{a priori}, they are not well suited for online settings where newly created experimental units must be assigned in real-time and assignment may depend on unit characteristics not available in advance.

The prevalence of randomized experiments in the Internet industry and the tools developed there are only partially represented in the scholarly literature. Mao et al. \cite{mao2012turkserver} created experimental frameworks for crowdsourcing sites such as Amazon Mechanical Turk. Several papers by Kohavi et al. present recommendations on how to implement and instrument experiments \cite{kohavi2009controlled}, as well as common pitfalls in analyzing experiments \cite{crook2009seven,kohavi2012trustworthy}.

Existing experimentation tools include associating experiments with ``layers'' (at Google and Microsoft \cite{kohavi2012trustworthy,tang2010overlapping}) or ``universes'' (at Facebook), such that all conditions in the same layer are mutually exclusive. Some tools (e.g., Google Analytics, Adobe Target) include mechanisms for associating condition identifiers with configuration information (e.g., dictionaries of parameters and their values).

While the types of experiments we focus on in this paper are designed to inform product decision-making, other experiments are run simply to optimize a single outcome variable.  For example, another active area of development focuses on implementing heuristics for optimizing stochastic functions (e.g., multi-armed bandit optimization) \cite{li2010contextual,scott2010modern}.


\begin{figure*}[!h]
\begin{center}
\includegraphics[width=0.93\textwidth]{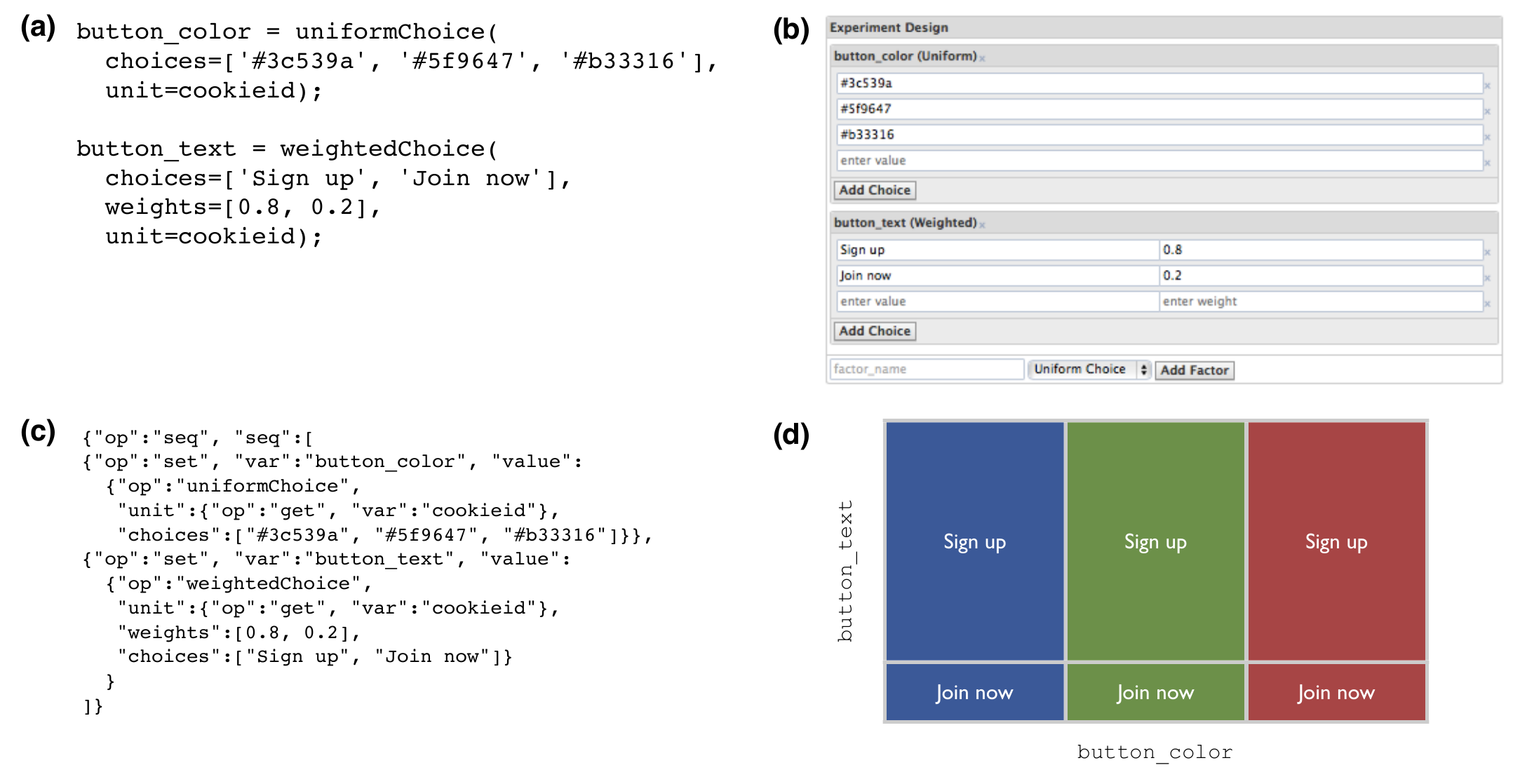}
\caption{A factorial experiment in PlanOut. (a) PlanOut language script (b) a graphical interface for specifying simple PlanOut experiments (c) a JSON representation of serialized PlanOut code (d) an illustration of the proportion of \ic{cookieids} allocated to each parameterization. Note that because we use \ic{weightedChoice()} to assign \ic{button\_text}, more cookies are assigned to ``Sign up'' than `'Join Now''.}
\label{fig:interfaces}
\end{center}
\end{figure*}
\section{The PlanOut language}
\label{sec:planout}
The PlanOut language separates experimental design from application logic and focuses experimenters on the core aspect of an experiment: how \emph{units} (e.g., users, items, cookies) are randomly assigned to conditions, as defined by parameters (e.g., settings for user interface elements, references to ranking algorithms). PlanOut promotes a mental model where every aspect of the site is parameterizable, and experiments are a way of evaluating user experiences defined by those parameters.

This approach encourages experimenters to decompose large changes into smaller components that can be manipulated independently. In doing so, experimenters are better equipped to attribute changes in user behavior to specific features and thus inform design decisions. Decomposition also allows experimenters to more easily iterate on experiments so that some features remain fixed while others change.

Experimenters use PlanOut by writing a PlanOut script, which may then be executed via an API for each unit (e.g., user or user-story combination). Each script indicates which inputs are used for assignment, how random assignment should occur, and the names of parameters that can be accessed via the API and logged. PlanOut scripts are executed sequentially.
In later sections, we introduce a number of scripts for both standard and complex experiments.

From a systems perspective, PlanOut is a way of serializing experiment definitions (e.g., as JSON) so they can be easily stored and executed on multiple platforms, such as layers of backend and frontend services, and mobile devices.
Serialized PlanOut code (Figure~\ref{fig:interfaces}c) can be generated and edited through a domain-specific language (DSL) (Figure~\ref{fig:interfaces}a) or graphical user interfaces (Figure~\ref{fig:interfaces}b).  The DSL is presented in the remaining sections, but many of these examples could alternatively be formulated through the use of graphical interfaces.

The PlanOut DSL and its syntax are minimal. The primary contribution of the language is in providing a parsimonious set of operations for thinking about, designing, and implementing experiments. Because of this, we spend little time discussing the language itself or its built-in operators. A complete list of operators can be found in the documentation for the reference implementation of PlanOut.\footnote{An open source implementation of a PlanOut interpreter and API is available at \ic{https://github.com/facebook/planout}.}
\subsection{Functionality}
We begin our discussion of PlanOut by giving several illustrative examples of how scripts and operators work. We first describe a simple A/B test and show how it can be generalized into factorial designs. Then we consider how experimental designs that involve multiple types of units in the randomization of a user interface element can be used to estimate different effects. Finally, we discuss conditional evaluation (e.g., for pre-stratification), and other operators.
\subsubsection{A/B test}
The most common type of experiment involves uniform random selection --- for example randomly assigning users to one of several button colors or text options. This can be accomplished via the \ic{uniformChoice} operator,

\code{
\tab{0.2}button\_color = uniformChoice(
\ntab{0.4}choices=['\#3c539a', '\#5f9647', '\#b33316'],
\ntab{0.4}unit=cookieid);
}

\noindent Here, each \ic{cookieid} is assigned deterministically to one of three possible button colors. In application code, the experimenter will later be able to evaluate this PlanOut script for a particular \ic{cookieid} (e.g., in the case of a Web-based sign-up form), and retrieve the runtime value through the variable name \ic{button\_color}.

\subsubsection{Multifactor experiment}
Creating a full factorial experiment means setting multiple variables that are evaluated independently. For example, suppose we wanted to manipulate not just the color of a button but also its text.  The script for this experiment is given in Figure~\ref{fig:interfaces}a, and includes two operators, a \ic{uniformChoice} and a \ic{weightedChoice}.  We use \ic{weightedChoice} to assign (on average) 80\% of the cookies to have the button text  ``Sign up'', and 20\% to have the text ``Join now''.  Setting these two parameters as shown in Figure~\ref{fig:interfaces} generates $2 \times 3 = 6$ conditions, whose proportions are summarized in Figure~\ref{fig:interfaces}d.

\subsubsection{Conditional execution}
Many experiments cannot be described through fully factorial designs, e.g., in the case where some values of one parameter may only be valid when another parameter is set to a particular value, or assignment probabilities are dependent on another variable. PlanOut thus includes operators for conditional control flow, such as \ic{if} / \ic{else}, boolean operations (i.e., \ic{and}, \ic{or}, \ic{not}), comparison operations (e.g., \ic{==}, \ic{>=}) and array indexing.

Consider a scenario where we wish to control the population of users receiving a new translation feature so that a higher proportion of US users receive the feature. To accomplish this, one could pass in both a \ic{userid} and \ic{country} to the PlanOut interpreter, and use conditional logic,

\code{
\tab{0.2}if (country == `US') \{
\ntab{0.4}has\_translate = bernoulliTrial(p=0.2, unit=userid);
\ntab{0.2}\} else \{
\ntab{0.4}has\_translate = bernoulliTrial(p=0.05, unit=userid);
\ntab{0.2}\}
}
\noindent or alternatively, via array indexing,

\code{
\tab{0.2}strata\_p = [0.05, 0.2];
\ntab{0.2}has\_translate = bernoulliTrial(
\ntab{0.4}p=strata\_p[country == `US'],
\ntab{0.4}unit=userid);
}

\noindent Here, arrays are zero-indexed and true/false evaluate to \ic{1} and \ic{0}, respectively.

\subsubsection{Experiments with multiple and nested units}\label{sec:multiple_units}
Many effects are better understood through randomization of units other than the user. For instance, while most standard A/B tests are between-subjects designs, where users are randomly assigned to different experiences, some effects may be more precisely estimated through a within-subjects design. These experiments can be implemented by transitioning from a single experimental unit (e.g., \ic{viewerid}) to tuples of units.

\begin{figure}[tb]
\begin{center}
\centering
\subfloat[]{
\hspace{2pt}
\includegraphics[width=0.95\columnwidth]{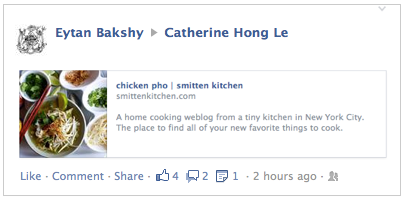}
\hspace{2pt}
}\\
\subfloat[]{
\hspace{2pt}
\includegraphics[width=0.95\columnwidth]{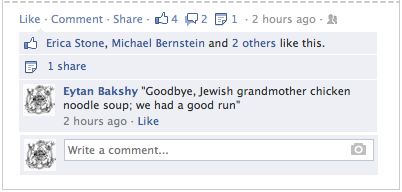}
\hspace{2pt}
}
\caption{(a) A News Feed story whose comment box is collapsed and (b) an expanded comment box. Different causal effects can be estimated by randomizing the state of the comment box over different experimental units (e.g., source users, viewers, stories). }
\label{fig:ufi}
\end{center}
\end{figure}

Consider an experiment that manipulates whether a story in users' News Feed has its comment box collapsed (Figure~\ref{fig:ufi}a) or expanded (Figure~\ref{fig:ufi}b). If an experimenter wanted to assign 5\% of all News Feed items to have a collapsed comment box, so that users must click to see comments attached to a story, one could define such an experiment by:

\code{
\tab{0.2}collapse\_story = bernoulliTrial(p=0.05,
\ntab{0.4}unit=[viewerid, storyid]);
}

\noindent The \ic{bernoulliTrial} operator returns \ic{1} with probability \ic{p}, and \ic{0} otherwise. By making \ic{unit} a tuple, \ic{[viewerid, storyid]}, one achieves a fully randomized within-subjects design, where each user sees, in expectation, an independent 5\% of posts with collapsed comment boxes. This type of design may be used to estimate the effect of collapsing individual stories on individual viewers' responses to that story (e.g., likes, comments).

\pagebreak
Other ways of assigning units can be used to estimate different quantities.  If one were to instead assign \ic{viewerid}s to conditions, 5\% of  users would see all stories collapsed, which could have large effects on how viewers interact with all stories and how many stories they would consume.  Had \ic{storyid} been the randomized unit, particular stories would be collapsed or not collapsed for all viewers. Because viewers are expected to comment at a higher rate when the box is expanded, randomizing over just \ic{storyid} can produce herding effects. These differences result from interference or ``spillovers'' across conditions \cite{aronow_estimating_2011,rubin_statistics_1986}, and highlights how supporting multiple experimental units can be useful for evaluating how user interface elements affect complex user dynamics.

\subsubsection{Extensions}
PlanOut is extensible.  If an assignment procedure cannot be implemented using built-in operators, developers may write custom operators in a native language (e.g., PHP or Python), including those that integrate with other services.  PlanOut experiments at Facebook often use custom operators that interface with gating infrastructure, graph cluster randomization~\cite{ugander_graph_2013}, and other experimentation systems.  Classes for random assignment are also extensible, so that procedures can be easily implemented using the hashing methods described below.

\subsection{Random assignment implementation}\label{sec:hashing}
Many operators involve generating pseudo-random numbers deterministically based on input data (e.g., a user generally should be assigned to a the same button color each time they load a particular page). A sound assignment procedure maps experimental units to parameters in a way that is deterministic, as good as random, and unless by design, independent of other parameter values from the same or other experiments.  And because experiments can be linked across multiple service layers (e.g., ranking \emph{and} user interfaces), it is important that any pseudo-random operation can be kept consistent across loosely-coupled services that may be written in different languages.

The PlanOut interpreter implements procedures that automatically fulfill these requirements.  Rather than using a pseudo-random number generator, or directly hashing units into numbers, the interpreter ``salts'' inputs to a hash function so that each assignment (unless otherwise specified) is independent of other assignments, both within and across experiments. Because the procedure based off of standard hashing functions (i.e., SHA1), it is deterministic and platform-independent.  At a low level, this is done by prepending each unit with a unique experiment identifier and variable-specific salt. Thus, the hash used to assign a variable such as \ic{button\_color} is not just the input ID (e.g., \ic{42}), but instead, e.g., \ic{user\_signup.my\_exp.button\_color.42}, where \ic{user\_signup} is the namespace of the experiment and \ic{my\_exp} is the identifier of the particular experiment (namespaces and experiments are more specifically defined in following sections).

\section{Example experiments}
The preceding examples show how PlanOut is a low-threshold language for implementing basic experiments.  In this section we demonstrate that PlanOut also has a high ceiling, in that complex, scientific experiments can be concisely specified in only a few lines of code.

\subsection{Examples from prior research}
We begin by demonstrating how one could implement two published experiments from the social computing literature.

\subsubsection{Experimenting with goal-setting}
In an influential application of social psychological theory to online systems, Beenen et al. \cite{beenen2004using} experimented with strategies for encouraging users to contribute ratings to the movie recommendation service and online community MovieLens. In Study 2, they randomly assigned users to an email that sets a goal for users to rate movies. Users were either identified as being part of a group, and having a group-level goal, or having an individual goal. The goal was either specific or a ``do your best'' goal; if specific, it was a number of movies to be rated in a week scaled by the size of the group. This experiment could be implemented as:

\code{
\tab{0.2}group\_size = uniformChoice(choices=[1, 10],
\ntab{0.4}unit=userid);
\ntab{0.2}specific\_goal = bernoulliTrial(p=0.8, unit=userid);
\ntab{0.2}if (specific\_goal) \{
\ntab{0.4}ratings\_per\_user\_goal = uniformChoice(
\ntab{0.6}choices=[8, 16, 32, 64], unit=userid);
\ntab{0.4}ratings\_goal = group\_size * ratings\_per\_user\_goal;
\ntab{0.2}\}
}

\noindent This experiment could then be analyzed in terms of the per-person specific goal, as in Beenen et al. \cite{beenen2004using}. There are multiple other ways to implement this experiment, some of which correspond to different choices about how to split logic between PlanOut and the application code. For example, the actual text used in the emails could be constructed in the PlanOut code, while the implementation above follows from the judgement that it would be better to do so in the application logic.
Such choices can also depend on other available tools, such as tools for automatically creating translation tasks for new strings used in an online service.

\subsubsection{A social cues experiment with complex inputs}
\label{sec:social_cues}
\begin{figure}
\begin{center}
\centering
\subfloat[]{
\hspace{1pt}
\includegraphics[width=0.47\columnwidth]{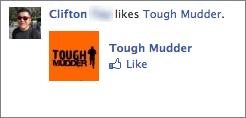}
\hspace{1pt}
}
\subfloat[]{
\hspace{1pt}
\includegraphics[width=0.47\columnwidth]{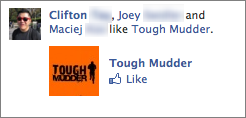}
\hspace{1pt}
}
\caption{A within-subjects experimental design that deterministically randomizes social cues presented to users. Example from Bakshy et al. [3] when (a) 1 of 3 and (b) 3 of 3 cues are shown.}
\label{fig:social_cues}
\end{center}
\end{figure}

Consider an experiment on the effects of placing social cues alongside advertisements from Bakshy et al. \cite{bakshy2012social}. A small percentage of user segments were allocated to this experiment (see Section \ref{sec:namespaces}); for these users, some social cues were removed from ads. For instance, if a user in the experiment had three friends that ``like'' a particular Facebook page being advertised, then this user would be randomly assigned to see one, two, or three friends associated with the page (Figure~\ref{fig:social_cues}).
This experiment can be written as follows:

\code{
\tab{0.2}num\_cues = randomInteger(
\ntab{0.4}min=1, max=min(length(liking\_friends), 3),
\ntab{0.4}unit=[userid, pageid]);
\ntab{0.2}friends\_shown = sample(
\ntab{0.4}choices=liking\_friends, draws=num\_cues,
\ntab{0.4}unit=[userid, pageid]);
}

\noindent The input data to the PlanOut experiment would be \ic{userid}, \ic{pageid}, and \ic{liking\_friends}, an array of friends associated with the page.  The script specifies that each user--page pair is randomly assigned to some number of cues, \ic{num\_cues}, between one and the maximum number of displayable cues (i.e., no more than three, but no greater than the number of friends eligible to be displayed alongside the ad).  That number is then used to randomly sample \ic{num\_cues} draws from \ic{liking\_friends}. That is, the script determines both the number of cues to display and the specific array of friends to display.

\subsection{Experiments deployed using PlanOut}
\label{sec:deployed_examples}
The following represent a sample of experiments that have been designed and deployed using PlanOut at Facebook.

\subsubsection{Voter turnout experiment}
\label{sec:gotv}
In an extension and replication of prior experiments with voter turnout \cite{bond_massive_2012,bryan_motivating_2011}, an experiment assigned all voting-aged US Facebook users to encouragements to vote in the 2012 Presidential Election. This experiment involved assigning users to both a banner at the top of the screen and eligibility for seeing social stories about friends' self-reported voting behavior in News Feed. We show a subset of the parameters set by this experiment.

\code{
\tab{0.2}has\_banner = bernoulliTrial(p=0.97, unit=userid);
\ntab{0.2}cond\_probs = [0.5, 0.98];
\ntab{0.2}has\_feed\_stories = bernoulliTrial(
\ntab{0.4}p=cond\_probs[has\_banner],
\ntab{0.4}unit=userid);
\ntab{0.2}button\_text = uniformChoice(
\ntab{0.4}choices=[``I'm a voter'', ``I'm voting''],
\ntab{0.4}unit=userid);
}

\noindent
We can see that \ic{has\_banner} is \ic{1} for 97\% of users.  Then, we define \ic{cond\_probs} to be the conditional probability that one would show feed stories given that \ic{has\_banner} is either \ic{0} or \ic{1}.  We assign \ic{has\_feed\_stories} using a \ic{bernoulliTrial()} with \ic{p=cond\_probs[has\_banner]}, so that those with the banner have a high chance of also being able to see the feed stories, and those without a banner have an equal probability of being able to see or not see the feed stories.  Finally, the button text for the call to action in the banner is subject to experimental manipulation (provided that the user is in the \ic{has\_banner=1} condition).  Analyses can then examine effects of the banner, effects of social stories about voting, and its interaction with verb/noun phrasing~\cite{bryan_motivating_2011}.

\subsubsection{Continuous-treatment encouragement design} 
Encouragement designs \cite{holland_causal_1988} randomize an inducement to a behavior of interest so as to evaluate the inducement or study the behavior's downstream effects. In online services, it is common to encourage users to engage with a particular entity, user, or piece of content. For instance, if having more ties on Facebook is hypothesized to increase long-term engagement, one could establish a causal relationship by randomizing whether or not some users receive recommendations for additional friends.

Evidence suggests that users who receive more feedback on Facebook are more likely to become engaged with the site~\cite{burke2009feed}. If there is a (forward) causal relationship between these variables, then changes to the site that affect how much feedback users receive can in turn affect user engagement and content production.

The following experiment examines this hypothesized effect by randomizing encouragements for friends to engage with a source user's content. It also illustrates random assignment involving multiple units. As mentioned in Section~\ref{sec:multiple_units}, expanding or collapsing News Feed stories can affect engagement with stories. The script below randomly assigns each source user to a proportion, such that on average, that proportion of the source's friends see a collapsed comment box when stories they produce appear in News Feed.

\code{
\tab{0.2}prob\_collapse = randomFloat(min=0.0, max=1.0,
\ntab{0.4}unit=sourceid);
\ntab{0.2}collapse = bernoulliTrial(p=prob\_collapse,
\ntab{0.4}unit=[storyid, viewerid]);
}

\noindent
Each source user is assigned to a probability \ic{prob\_collapse} in $[0, 1]$.  Then, each story--viewer pair is assigned to have a collapsed comment box with probability \ic{prob\_collapse}.  To carry out this assignment, we invoke the PlanOut script from the part of News Feed rendering logic that determines whether stories' comment boxes should be expanded or collapsed, using \ic{sourceid}, \ic{storyid}, and \ic{viewerid} as inputs (more discussion of the application interface is covered in Section~\ref{sec:api}).

There a number of possible ways this experiment can be analyzed. First, one can identify the effect of modulating feedback encouragements on the total amount of feedback a user's stories get.  Second, we can test our original hypothesis that feedback causes users to engage more with the site (e.g., log in more often or produce more content). One can look at the effect of the assignment to different values of \ic{prob\_collapse} on users' engagement levels, or use an instrumental variables analysis \cite{holland_causal_1988,morgan2007counterfactuals}, which combines estimates of effects of the encouragement on feedback received and engagement, to estimate the effect of feedback on users' engagement.

\section{Running experiments}
\label{sec:running}
We have discussed ways of designing and executing randomized assignment procedures, but have not described how experiments are managed, tested, and logged. Here we define an \emph{experiment} to refer to a PlanOut script combined with a target population for which that script was launched at a specific point in time.  From the perspective of the experimenter and logging infrastructure, different experiments are considered separately.

In the following subsections, we describe a broader technical context for running experiments.
This supporting infrastructure includes: a system for managing and logging experiments, including a segmentation layer which maps units to experiments, a launch system which provides default values for parameters not subject to experimental manipulation, an API for retrieving parameters, and a logging architecture which simplifies data analysis tasks.

\subsection{Namespace model of site parameters} \label{sec:namespaces}
Field experiments with Internet services frequently involve the manipulation of persistent parameters that are the subject of multiple experiments, whether conducted serially or in parallel. We use a namespace model of parameters to support these practices.

Experimentation is frequently iterative; as in scientific research, a single experiment is often not definitive and so requires follow-up experiments that manipulate the same parameters. A second experiment with a near-identical design may be used to more precisely estimate effects, or might include new variations suggested by the first's results or other design work. Continual redesigns and development might also change the effects of the parameters, thus motivating the need for additional experiments.

Likewise, multiple experiments manipulating the same aspects of a service are frequently run in parallel, sometimes by different teams with minimal explicit coordination. Two teams may manipulate (a) the same features of the same service, (b) independent features of the same service (e.g., font size and items per page), or (c) different layers of the same service (e.g., link colors and the ranking model which selects which items are to be displayed). In these cases, it is helpful to have an experimentation system that can keep track of and/or restrict which parameters are set by each experiment.
This requires that experimentation tools be cross-platform and can handle allocation of units to multiple experiments started at different times by different teams.

The solution to support these practices within the PlanOut framework is to use \emph{parameter namespaces} (or namespaces for short). This is a natural extension of thinking of the experimentation system as being how many parameter values are read in application code; often these parameters are an enduring part of the service, such that, over time, many experiments will set a particular parameter. Each namespace is centered around on a primary unit (e.g., users). A new experiment is created within a new or existing namespace by allocating some portion of the population to a PlanOut script.\footnote{Any two experiments which set the same parameter must be mutually exclusive of one another (i.e., must assign parameter values only for disjoint sets of units). More generally, consider the graph of experiments in which two experiments are neighbors if they set any of the same parameters. It is then natural to require that all experiments in the same connected component to be mutually exclusive. This motivates the idea of using namespaces to group parameters that are expected to be set by experiments in the same connected component of this graph.}

\subsection{Experiment management}
Experiments can be managed as follows: for each namespace, hash each primary unit to one of a large number (e.g., 10,000) of \emph{segments}, and then allocate individual segments to experiments. This segment-to-experiment mapping may be maintained in a database or other data storage system.  Similar to segmentation systems discussed in prior work \cite{tang2010overlapping}, when a new experiment is created, it is allocated a random set of segments.  These segments are deallocated once the experiment is complete.

\begin{figure}
\begin{center}
\includegraphics[width=1.0\columnwidth]{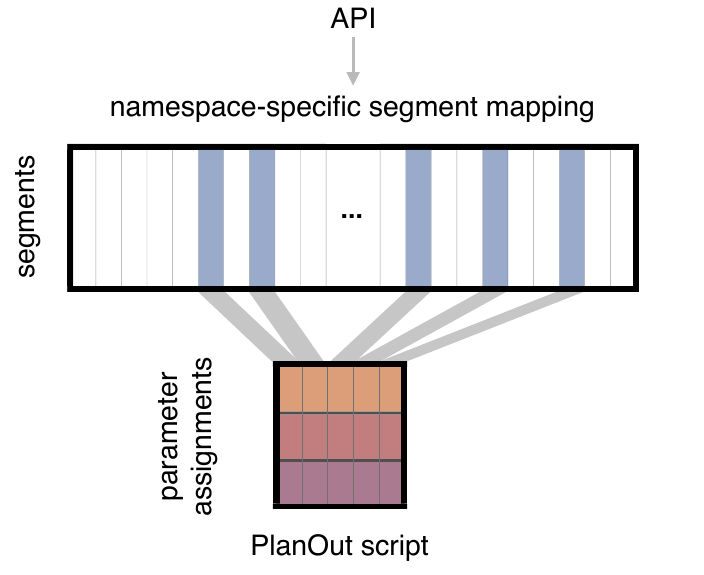}
\caption{Blocks of segments (e.g., multiple buckets of user ids) are assigned to experiments, which map experimental units to parameters in a way that is uncorrelated with the segmentation.}
\vspace{5pt}
\label{fig:segmentation}
\end{center}
\end{figure}

Each experiment's script makes no reference to these segments, such that random assignment to parameter values within each experiment is independent of the segmentation (Figure \ref{fig:segmentation}).\footnote{This corresponds to what Kohavi et al. \cite{kohavi2012trustworthy} call ``local randomization''. It requires that each experiment have its own dedicated control group, which they identify as the approach's lone disadvantage.} This feature is accomplished via the hashing method described in Section~\ref{sec:hashing}, and reduces the risk of carryover effects \cite{box2005statistics} that might occur if whole segments from one experiment were all assigned to the same parameterizations and subsequently reallocated to a different experiment~\cite{kohavi2012trustworthy,kohavi2009controlled}.

\begin{figure*}[!ht]
\begin{center}
\includegraphics[width=1\textwidth]{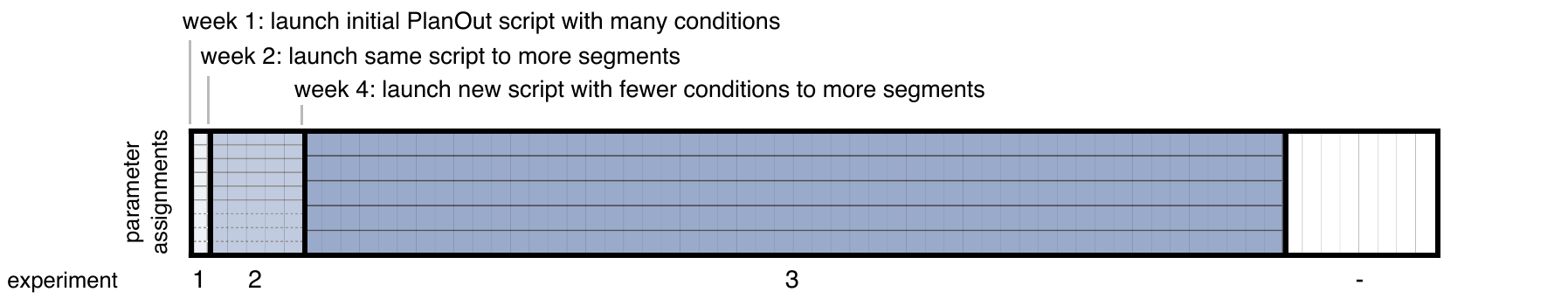}
\caption{An illustration of how namespaces are used to launch experiments. Segments (vertical regions) are randomly allocated to experiments; here we order the segments by which experiment they were allocated to. Segments not allocated to experiments use the default parameter values. Results from each experiment are generally analyzed separately. Horizontal lines distinguish conditions in each experiment; dotted lines indicate the conditions removed in experiment 3.}
\label{fig:iteration}
\end{center}
\end{figure*}

Namespaces can be represented in experimentation management tools as a dashboard of currently running experiments along with a listing of parameters for that namespace.
When an experiment is complete, its segments can become available for future experiments. When iteratively experimenting, new versions of an experiment can be created and allocated new segments. For example, when experimenters need to increase precision by experimenting with more units, engineers can duplicate the experiment definition and allocate additional segments to the experiment. Increases in size are often coupled with changes to the experiment definition, e.g., adding new parameterizations similar to promising versions. Such iteration is generally preferable to modifying the existing experiment, which can frequently produce problems in analysis (see Section \ref{sec:Analysis}).

\subsubsection{Parameter defaults}\label{sec:defaults}
In some cases, all units will have a parameter set by an experiment. For example, an experimenter working with a new parameter in a new namespace may simply allocate all segments to an experiment. But in other cases, some units will not have a particular parameter set by any experiment. Then the value of the parameter used for those units can be set in some other way. Often, this would reflect the status quo and/or what values are currently believed to be optimal. This same value is likely assigned by other experiments (e.g., if they include ``control'' conditions).\footnote{From a statistical perspective, it may seem that not counting these units as part of a control group makes for inefficient analysis. This is sometimes true, but doing so would place additional requirements for standard analyses to be correct (e.g., requirements on the history of values assigned to units not in experiments). When these requirements are satisfied, an analyst could make use of that data as needed.
}

More technically, if a request for a parameter is made but that parameter is not set by an experiment (i.e., the unit is not assigned to an active experiment, or the unit is assigned to an active experiment that does not set the parameter), then some specified \emph{launch value} is used.  If this launch value is not specified, or the experimentation service fails, the default specified in the application code is used instead. Other extensions, such as having launch values vary depending on unit characteristics (e.g., country) may also be implemented in a straightforward way.

\subsubsection{Workflow for iterative experimentation}
We summarize how the tools presented here fit together by describing an iterative experiment (Figure~\ref{fig:iteration}).  First, we created a namespace for a particular user interface, implemented front-end code to retrieve parameter values from that namespace immediately before rendering UI elements, and set the default launch values to settings that had previously been hard-coded in PHP. In our case, logging for the outcomes of interest was already instrumented, so no additional instrumentation was needed.

Then, we scripted an initial PlanOut-based experiment and launched it to a small set of users (experiment 1 in Figure~\ref{fig:iteration}).   After one week of observing that the experiment did not cause statistically significant decreases in key metrics, this same script was launched to roughly 8 times the number of users (experiment 2).  Results were initially analyzed using internal tools, and then in greater detail using \ic{R}. Because the experiment was expected to have long-term effects, results from each experiment were analyzed separately even though they used the same script.  We found that the primary outcomes were clearly worse under parameterizations involving a particular parameter value, and that higher values of a second parameter appeared to increase one outcome at the expense of another outcome.

Based on these results, we created a new PlanOut script that did not include clearly suboptimal parameterizations, and extended the range of the parameter we hypothesized to represent an important tradeoff.  This new script was launched in a third experiment to an additional segment of users (experiment 3 in Figure~\ref{fig:iteration}), and analyzed for several weeks. 

After considering longer-term results from the three experiments, we decided on a parameterization to use as a default for all users.  We de-allocated all segments, set a new default parameterization, and created a new, smaller experiment that assigned users to the new and original parameterizations with equal probability. This fourth experiment, commonly referred to as a ``backtest'', is used to evaluate the efficacy of the launch decision after a long period of time.

While this type of backtest is easy to implement and avoids the potential for downstream errors in analysis, there are a number of possible ways we could have run the backtest.  If a prior experiment (e.g., experiment 3) had reasonable power for the comparison between the old and new defaults, we could continue to run that experiment and disable all other parameterizations.  Users assigned to disabled parameterizations would take on the new default parameterization, but would not be used in the subsequent analysis of the experiment. This approach can be particularly attractive if the experiment is expected to have time-varying effects, or if one wanted to minimize changes to individual users' experiences.  A third option would involve running a new experiment in the unallocated segments, if there are enough such segments after experiments 1--3. 

\subsection{Integration with application code}\subsubsection{Application programming interface}\label{sec:api}
At runtime, application code needs to retrieve parameter values within a namespace for particular units.  This can be done by constructing an object which interfaces with management and logging components.  For example, retrieving the parameter \ic{collapse\_story} associated with a particular viewer-story pair in the \ic{comment\_box} namespace might be invoked by instantiating an experiment object for the input units, and request the needed parameters:

\code{
\tab{0.2}exp = getExp(`comment\_box',
\ntab{0.4}\{`viewerid': vid, `storyid': sty\})
\ntab{0.2}collapse\_story = exp.get(`collapse\_story')
}

\noindent
The management system would then map the units to a segment within the \ic{comment\_box} namespace, which gets mapped to an experiment and its respective PlanOut script. The script is executed with the input data, and if  the script that sets a parameter requested by \ic{get()}, the value is returned and the event is logged. If the requested parameter is not set, then the parameter default (described in Section~\ref{sec:defaults}) is used.
Because assignment procedures bear some computational cost, and are generally deterministic, parameter assignments can be cached.

\subsubsection{Testing experiments and parameter values}
When designing experiments, engineers often need to be able to test a service under a range of parameter values.  While PlanOut scripts can be difficult to interact with directly (as they are not written in a native language, like PHP), they can still be tested and debugged in situ with a small amount of additional infrastructure. In particular, by providing developers with a way to override or ``freeze'' parameters so that they maintain a prespecified value throughout a PlanOut script's execution, one can test assignment to conditions even if few units (or combinations of units) are assigned to them, without modifying any application or PlanOut code.

This functionality can be surfaced to Web developers via URL query parameters. Freezing the \ic{has\_feed\_stories} parameter to \ic{1} in the voter turnout experiment described in Section~\ref{sec:gotv} (running within the \ic{vote2012} namespace) may then be accomplished by accessing a URL like:

\code{http:\//\//...php?ns\_vote2012=has\_feed\_stories:1}

Freezing also allows one to test downstream effects of different inputs. Overriding \ic{userid} or \ic{has\_bannner} may in turn change whether feed stories are shown.  Combinations of parameters can also be frozen by specifying a list of parameters to be set, e.g. \ic{has\_banner:1,has\_feed\_stories:0}. Overrides for mobile applications or backend services may alternatively be set through server-side management tools.
\subsection{Logging}
Logging occurs automatically when \ic{get()} is called, so that there is a record of the exposure of units to the experiment.  By default, the namespace, the experiment name, all input data, and variables set by the PlanOut script are logged.  This type of \emph{exposure logging} has a number of benefits, including simplifying downstream data analysis and increasing statistical power by distinguishing between units that may have been affected by assignment and those that are known to be unaffected. (For example, many users who are assigned to be in an experiment may not actually arrive at the part of the site that triggers the manipulation, and thus their outcome data should not be included in analysis in the normal way.) As with experimental assignment, caching of prior exposures help reduce load on experimental infrastructure.  It is also sometimes desirable to log auxiliary information not related to the assignment, including user characteristics or events (i.e., ``conversions'').  This might be done through a separate method (e.g., \ic{log()}) in the experiment object. Exposure logging, combined with the management system, prevents a number of common pitfalls we have observed at Facebook. These benefits are discussed in the following section. 

\section{Analyzing Experiments} \label{sec:Analysis}
While domain-specific aims and especially complex experimental designs will often require custom analysis, much analysis of online experiments can be automated in support of their routine and valid use in decision making. This eliminates common sources of errors in analysis, makes results more directly comparable, and reduces the burden of running additional experiments. This kind of automation is easy to accomplish with PlanOut experiments because their scripts directly encode a representation of their parameters, values, and design. A complete description of accompanying systems for analyzing experiments is beyond the scope of this paper, but we discuss how the design of PlanOut interacts with common analyses, automated or not.

Logging only users who have received the treatment (versus analyzing the entire population who could could have potentially be exposed) improves statistical inference in two ways. First, it can substantially decrease the variance of estimated treatment effects when the number of users exposed to an experimental manipulation is small relative to the number who are assigned (e.g., in the case of a less commonly used or new feature).  This reduces risk of Type II errors, in which an experiment has an effect, but experimenters are unable to detect that effect.  Secondly, exposure logging focuses experimenters on a more relevant sub-population whose outcomes are plausibly affected by the treatment.

Explicit logging of labeled input units and assigned parameter values affords flexibility in terms of automated analysis. Many relatively simple designs can be fruitfully analyzed by computing summary statistics for outcomes of interest for each unique combination of parameter values. Since parameters are logged for each exposure, analyses of full or fractional factorial designs that make use of this structure can also be automated.  For example, questions about main effects of factors (e.g., ``does button color have any average effect?'') can be answered via an analysis of variance. Representation of the experiment in terms of parameters can also make it easier to automatically use this structure in estimating expected outcomes for each condition. For example, systems can fit penalized regression models with the main effects and all relevant interactions, thus ``borrowing strength'' across conditions that have the same values for some parameters \cite{gelman2005analysis}. These types of model-based inference also help reduce the risk of not being able to detect clinically significant changes.  We have found that when these forms of analyses are not possible, engineers and decision makers tend to avoid more complex experimental designs, like factorial designs, because they tend to be underpowered, even though they have a number of benefits for improving understanding and identifying optimal treatments.

\subsection{Analyzing iterative experiments}
Iteration on experiments with PlanOut occurs primarily through creating new experiments that are variants of previous experiments. By default, these experiments are then analyzed separately, which avoids several problems that can occur when attempting to pool data from before and after a change in experimental design. For example, adding additional users to an existing experiment but assigning them to new conditions means that these users are first exposed to their treatment more recently than other users; comparisons with other conditions can be biased by, e.g., novelty effects or cumulative treatment effects.

There are some cases where two similar experiments can be analyzed together.
For example, if two experiments' PlanOut scripts are identical but have different numbers of segments allocated to them and are started at different times, they can be pooled together to increase power, though this changes what is estimated to a weighted average of potentially time-varying effects.  This may result in underestimation or overestimation of treatment effects, and highlights the ways in which gradual product rollouts might be better represented as experiments. More sophisticated automatic selection of analyses that pool data across experiments remains an area for future work.

\subsection{Units of analysis}
Many online experiments use a small number of standard types of units, for which outcomes of interest may already be available in data repositories. For example, most experiments at Facebook involve random assignment of user IDs, and the standard desired analysis involves analysis of behaviors associated with user IDs. Other cases can be more complex. For example, an experiment may randomly assign users and advertisements, userid--advertisementid pairs, to parameter values, but it may be necessary for an analysis to account for dependence in multiple observations of the same user or ad to obtain correct confidence intervals and hypothesis tests \cite{bakshy2013bootstrap,cameron2006robust}. Inspection of a script can identify the units for which different parameters are randomized, which can be used in subsequent selection of methods for statistical inference.

\section{Discussion}
\label{sec:discussion}
Randomized field experiments are a widely used tool in the Internet industry to inform decision-making about product design, user interfaces, ranking and recommendation systems, performance engineering, and more. Effective use of these experiments for understanding user behavior and choosing among product designs can be aided by new experimentation tools. We developed PlanOut to support such scalable, randomized parameterization of the user experience. Our goal has been to motivate the design of PlanOut using our experiences as experimenters and by demonstrating its ability to specify experimental designs, both simple and complex, from our work and the literature.

One aim of conceptualizing experiments in terms of parameters and enabling more complex experimental designs is that online experiments can be more effectively used for understanding causal mechanisms and investigating general design principles, rather than simply choosing among already built alternatives. That is, PlanOut aims to support uses of randomized experiments more familiar in the sciences than in the Internet industry. For example, the primary purpose of the social cues experiment described in Section~\ref{sec:social_cues} is not to decide whether it is better to show fewer social cues alongside ads (doing so was expected to and did reduce desired behaviors),
but to estimate quantities that are useful for understanding an existing service, allocating design and engineering resources, and anticipating effects of future changes.

In addition to being a means for deploying Internet-scale field experiments, we have found PlanOut to be useful aid for describing and collaborating on the design of complex experiments, well before they are deployed. We hope others will also find the notation to be a clear way describe their experiments, whether in face-to-face settings or documentation of published research. The PlanOut language itself may also be applicable to other types of experiments, such as those conducted on Amazon Mechanical Turk \cite{mason2012conducting}.

There are important limitations to online experiments in general and PlanOut in particular. As others have argued, randomized experiments cannot effectively replace all other methods for learning from current and potential users \cite{nielson_putting_2005} and anticipating effects of future interventions \cite{deaton_instruments_2010}. Most notably, random assignment of users to a new version of a service requires that that version is built and of sufficient quality. Nielson \cite{nielson_putting_2005} additionally argues that A/B tests encourage short-term focus and do not lead to behavioral insights. While this is perhaps a fair critique of many widespread experimentation practices, PlanOut is designed to run experiments that lead to behavioral insights, modeling, and long-term learning.

Even though field experiments are the gold standard for causal inference, their results are also subject to limitations. Because the underlying effects can be heterogeneous and dynamic, results of a field experiment from one time and one population may not generalize to new times and populations \cite{bareinboim2012transportability,manzi2012uncontrolled,watts2011everything}.  One hope that we have for PlanOut is that it encourages more sophisticated behavioral experiments that allow estimation of parameters that are more likely to generalize to future interfaces. PlanOut and the associated infrastructure also make it easy to replicate prior experiments. Finally, standard experimental designs and analyses do not account for one unit's outcomes being affected by the assignment of other units (e.g., because of for peer influence and network effects) \cite{aronow_estimating_2011,rubin_statistics_1986}.  In the presence of such interference, user behavior can substantially change post-launch behaviors as connected users interact with one another \cite{ugander_graph_2013}.

PlanOut has more specific limitations with respect to designs where one unit's assignment depends on the assignment of a large number of other units. Random assignment schemes that involve optimizing global characteristics of the experimental design are thus more difficult to implement directly using built-in operators. This includes pre-stratified or block-randomized designs \cite{box2005statistics,gerber2012field} that use sampling without replacement in a prior, offline assignment procedure, but in online experiments these designs usually offer minimal precision gains.\footnote{The differences between the variance of a difference in means from a pre-stratified design and a post-stratified estimator with a unstratified design is of order $1 / n^{2}$ \cite{miratrix_adjusting_2013}. This difference is thus of little importance for large experiments.} Assignment schemes such as graph cluster randomization \cite{ugander_graph_2013}, which involves partitioning the social network of users, require offline computation. In such cases, PlanOut may simply be a useful framework for providing a consistent interface to accessing information about units (e.g., the results of the graph partitioning) that has been computed offline through a custom operator and then assigning parameter values based on that information.  Sequential experimentation techniques, such as multi-armed bandit heuristics \cite{scott2010modern} are another such example where custom operators are generally needed.

We have only briefly discussed how online experiments should be analyzed. From our experience, the availability of easy-to-use tools for routine analysis of experiments has been a major factor in the adoption of randomized experiments across product groups at Facebook; so the automation of analysis deserves further attention. This also suggests another area for future research: How should we best evaluate new tools for designing, running, and analyzing experiments? We have primarily done so by appealing to prior work, our own professional experiences, and by demonstrating the expressiveness of the PlanOut language. Deciding whether an experiment was ``successful'' or effective can depend on broader organizational context and hard-to-trace consequences as an experiment's results diffuse throughout an organization. Some of the most effective experiments directly inform decisions to set the parameters they manipulate, but other well-designed experiments can be effective through broader, longer-term influences on beliefs of designers, developers, scientists, and managers.

\section{Acknowledgements}
As described here, PlanOut is only a small piece of the broader set of experimentation tools created by our colleagues.  At Facebook, PlanOut runs as part of QuickExperiment, a framework developed by Breno Roberto and Wesley May.  The perspective we take on how experiments should be logged and managed is greatly influenced by previous tools at Facebook and conversations with Daniel Ting, Wojciech Galuba, and Wesley May. The design of PlanOut was influenced by conversations with John Fremlin, Brian Karrer, Cameron Marlow, Itamar Rosenn, and those already mentioned.  Finally, we would like to thank Brian Davison, Ren\'e Kizilcec, Winter Mason, Solomon Messing, Daniel Ting, and John Myles White for their comments on this paper.
Ren\'e Kizilcec built the PlanOut GUI in Figure \ref{fig:interfaces}(c). John Fremlin built the PlanOut DSL to JSON compiler. 

%
%
%
%
%
\balance
\bibliographystyle{acm-sigchi}
\bibliography{planout}
\end{document}